\documentclass[12pt]{article}

\usepackage{amsmath,amsfonts,amssymb,color,epsfig,graphics,graphicx,latexsym,revsymb,theorem,url,verbatim}

\usepackage[cm]{fullpage}

\newtheorem{definition}{Definition}
\newtheorem{proposition}[definition]{Proposition}
\newtheorem{lemma}[definition]{Lemma}

\newtheorem{theorem}[definition]{Theorem}
\newtheorem{corollary}[definition]{Corollary}
\newtheorem{conjecture}[definition]{Conjecture}

\newtheorem{remark}[definition]{Remark}
\newtheorem{example}[definition]{Example}
\newtheorem{question}[definition]{Question}

\def\squareforqed{\hbox{\rlap{$\sqcap$}$\sqcup$}}
\def\qed{\ifmmode\squareforqed\else{\unskip\nobreak\hfil
\penalty50\hskip1em\null\nobreak\hfil\squareforqed
\parfillskip=0pt\finalhyphendemerits=0\endgraf}\fi}
\def\endenv{\ifmmode\;\else{\unskip\nobreak\hfil
\penalty50\hskip1em\null\nobreak\hfil\;
\parfillskip=0pt\finalhyphendemerits=0\endgraf}\fi}
\newenvironment{proof}{\noindent \textbf{{Proof.~} }}{\qed}
\def\Dbar{\leavevmode\lower.6ex\hbox to 0pt
{\hskip-.23ex\accent"16\hss}D}
\makeatletter
\def\url@leostyle{%
  \@ifundefined{selectfont}{\def\UrlFont{\sf}}{\def\UrlFont{\small\ttfamily}}}
\makeatother
\def\bcj{\begin{conjecture}}
\def\ecj{\end{conjecture}}
\def\bcr{\begin{corollary}}
\def\ecr{\end{corollary}}
\def\bd{\begin{definition}}
\def\ed{\end{definition}}
\def\bea{\begin{eqnarray}}
\def\eea{\end{eqnarray}}
\def\bem{\begin{enumerate}}
\def\eem{\end{enumerate}}
\def\bex{\begin{example}}
\def\eex{\end{example}}
\def\bim{\begin{itemize}}
\def\eim{\end{itemize}}
\def\bl{\begin{lemma}}
\def\el{\end{lemma}}
\def\bpf{\begin{proof}}
\def\epf{\end{proof}}
\def\bpp{\begin{proposition}}
\def\epp{\end{proposition}}
\def\bqu{\begin{question}}
\def\equ{\end{question}}
\def\br{\begin{remark}}
\def\er{\end{remark}}
\def\bt{\begin{theorem}}
\def\et{\end{theorem}}

\def\btb{\begin{tabular}}
\def\etb{\end{tabular}}

\newcommand{\nc}{\newcommand}

\def\a{\alpha}

\def\p{\pi}

\def\ph{\varphi}

\def\ps{\psi}
\def\o{\omega}

 \nc{\bA}{{\bf A}} \nc{\bB}{{\bf B}}
 \nc{\bC}{{\mathbb{C}}}
 \nc{\bD}{{\bf D}} \nc{\bE}{{\bf E}} \nc{\bF}{{\bf F}}
 \nc{\bG}{{\bf G}} \nc{\bH}{{\bf H}} \nc{\bI}{{\bf I}}
 \nc{\bJ}{{\bf J}} \nc{\bK}{{\bf K}} \nc{\bL}{{\bf L}}
 \nc{\bM}{{\bf M}} \nc{\bN}{{\bf N}} \nc{\bO}{{\bf O}}
 \nc{\bP}{{\bf P}} \nc{\bQ}{{\bf Q}} \nc{\bR}{{\bf R}}
 \nc{\bS}{{\bf S}} \nc{\bT}{{\bf T}} \nc{\bU}{{\bf U}}
 \nc{\bV}{{\bf V}} \nc{\bW}{{\bf W}} \nc{\bX}{{\bf X}}
 \nc{\bZ}{{\bf Z}}

\nc{\cA}{{\cal A}} \nc{\cB}{{\cal B}} \nc{\cC}{{\cal C}}
\nc{\cD}{{\cal D}} \nc{\cE}{{\cal E}} \nc{\cF}{{\cal F}}
\nc{\cG}{{\cal G}} \nc{\cH}{{\cal H}} \nc{\cI}{{\cal I}}
\nc{\cJ}{{\cal J}} \nc{\cK}{{\cal K}} \nc{\cL}{{\cal L}}
\nc{\cM}{{\cal M}} \nc{\cN}{{\cal N}} \nc{\cO}{{\cal O}}
\nc{\cP}{{\cal P}} \nc{\cQ}{{\cal Q}} \nc{\cR}{{\cal R}}
\nc{\cS}{{\cal S}} \nc{\cT}{{\cal T}} \nc{\cU}{{\cal U}}
\nc{\cV}{{\cal V}} \nc{\cW}{{\cal W}} \nc{\cX}{{\cal X}}
\nc{\cZ}{{\cal Z}}

\nc{\hA}{{\hat{A}}} \nc{\hB}{{\hat{B}}} \nc{\hC}{{\hat{C}}}
\nc{\hD}{{\hat{D}}} \nc{\hE}{{\hat{E}}} \nc{\hF}{{\hat{F}}}
\nc{\hG}{{\hat{G}}} \nc{\hH}{{\hat{H}}} \nc{\hI}{{\hat{I}}}
\nc{\hJ}{{\hat{J}}} \nc{\hK}{{\hat{K}}} \nc{\hL}{{\hat{L}}}
\nc{\hM}{{\hat{M}}} \nc{\hN}{{\hat{N}}} \nc{\hO}{{\hat{O}}}
\nc{\hP}{{\hat{P}}} \nc{\hR}{{\hat{R}}} \nc{\hS}{{\hat{S}}}
\nc{\hT}{{\hat{T}}} \nc{\hU}{{\hat{U}}} \nc{\hV}{{\hat{V}}}
\nc{\hW}{{\hat{W}}} \nc{\hX}{{\hat{X}}} \nc{\hZ}{{\hat{Z}}}

\nc{\hn}{{\hat{n}}}

\def\dim{\mathop{\rm Dim}}

\def\lin{\mathop{\rm span}}

\def\max{\mathop{\rm max}}
\def\min{\mathop{\rm min}}

\def\sp{\mathop{\rm sp}}

\def\GL{{\mbox{\rm GL}}}

\def\bigox{\bigotimes}
\def\dg{\dagger}

\def\op{\oplus}
\def\ox{\otimes}

\def\su{\subset}
\def\sue{\subseteq}

\def\we{\wedge}

\newcommand{\ket}[1]{|#1\rangle}

\newcommand{\braket}[2]{\langle#1|#2\rangle}
\newcommand{\wetw}[2]{|#1\rangle\wedge|#2\rangle}

\newcommand{\abs}[1]{|#1|}

\def\Dbar{\leavevmode\lower.6ex\hbox to 0pt
{\hskip-.23ex\accent"16\hss}D}
\begin{document}
\title{Unextendible Product Basis for Fermionic Systems}

\author{Jianxin Chen$^{1,2}$, Lin Chen$^{2,3}$, Bei Zeng$^{1,2}$\\
\small ${}^{1}$ Department of Mathematics \&
Statistics, University of Guelph, Guelph, Ontario, Canada\\
\small ${}^{2}$ Institute for Quantum Computing, University of
Waterloo, Waterloo, Ontario, Canada\\
\small ${}^{3}$ Department of Pure Mathematics, University of Waterloo,
Waterloo, Ontario, Canada
}

\date{\today}

\maketitle

\begin{abstract}
We discuss the concept of unextendible product basis (UPB) and
generalized UPB for fermionic systems, using Slater determinants as
an analogue of product states, in the antisymmetric subspace $\we^ N
\bC^M$. We construct an explicit example of generalized fermionic
unextendible product basis (FUPB) of minimum cardinality $N(M-N)+1$
for any $N\ge2,M\ge4$. We also show that any bipartite antisymmetric
space $\we^ 2 \bC^M$ of codimension two is spanned by Slater
determinants, and the spaces of higher codimension may not be
spanned by Slater determinants. Furthermore, we construct an example
of complex FUPB of $N=2,M=4$ with minimum cardinality $5$. In contrast,
we show that a real FUPB does not exist for $N=2,M=4$ . Finally we
provide a systematic construction for FUPBs of higher dimensions
using FUPBs and UPBs of lower dimensions.
\end{abstract}

\section{Introduction}

Entanglement plays an essential role in many aspects in modern physics,
from foundations of quantum machines, mathematical physics,
to applied physics such as quantum information processing.
During the past decades, considerable progress
has been made to understand various aspects of entanglement~\cite{Hor09}. Majority
of these efforts were devoted to the understanding of entanglement
properties for indistinguishable particle systems given the obvious role they play
in quantum communication and computation. However, in some cases,
such as those quantum dot systems, identical particles also play an important
role in quantum information processing and their entanglement properties
needs different consideration given the restricted Hilbert space~\cite{SLM01}.

Identical particles are particles that cannot be distinguished from each
other, even in principle. Fermions are identical particles which
obey the Pauli exclusion principle, that is, no two identical
fermions may occupy the same quantum states simultaneous.
Mathematically, this means that for any many fermion system, the
quantum state $\ket{\psi}$ picks up a minus sign when interchanging
any two particles $i,j$, i.e.
$\text{SWAP}_{ij}\ket{\psi}=-\ket{\psi}$. In other words, the
fermionic wave function is anti-symmetric with respect to the
exchange of any two particles. For instance, for a two particle
state, $\ket{\psi}=\sum_{ij}a_{ij}\ket{i}\otimes\ket{j}$, then
$a_{ij}=-a_{ij}$, which implies $a_{ii}=0$, i.e. the Pauli exclusion
principle.

Entanglement in fermionic systems cannot be examined by the usual
way in the tensor product space. For instance, although a state as
$\frac{1}{\sqrt
2}(\ket{0}\otimes\ket{1}-\ket{1}\otimes\ket{0})$ looks
entangled in the tensor product space, this entanglement is not
accessible~\cite{SLM01}. Indeed, the labelling of particles and the notation of a
tensor product structure of the space of states are misleading
because the actually state space is the
antisymmetrized subspace, which is just a subspace of
the complete tensor product space.

It is now widely agreed that a $N$-fermion state which is obtained
solely by antisymmetrizing a computational basis state
$\ket{i_1}\otimes\ket{i_2}\cdots\otimes\ket{i_N}:=\ket{i_1,\cdots,i_N}$,
cannot be regarded as entangled~\cite{SLM01}, which can be written
as a determinant $\sum_{\sigma}\text{sgn}(\sigma)
\ket{i_{\sigma(1)},i_{\sigma(2)},\cdots,i_{\sigma(N)}}$, called a
Slater determinant. Here $\sigma$ is a permutation of
$(1,2\ldots,N)$. However, fermionic states that cannot be written as
a single Slater determinant of some
$\ket{i_1},\ket{i_2},\ldots,\ket{i_N}$ do contain useful
entanglement that can be used in practice to conduct nontrivial
quantum information procession and hence can be regarded as
entangled~\cite{sck01,esb02}.

During the past decade entanglement property in fermionic systems
has attracted significant amount of attention~\cite{Hor09,esb02,Zan02,ZX02,VC03,WV03,KL06}. A large amount of
effort was taken to look at analogies of fermionic entanglement to
entanglement in tensor product spaces, especially for low
dimensional systems. For $N=2$, it was shown that there exists a so
called `Slater decomposition' for any fermionic states, which is an
analogue of the Schmidt decomposition. Also,
for a single particle space of dimension $M=4$, several
interesting analogues with a two qubit system has been found, such
as `magic basis', `concurrence', and `dualisation'~\cite{esb02}.

Another interesting concept studied in entanglement theory is the
unextendible product basis (UPB) \cite{BDM+99}. For a multipartite
quantum system with Hilbert space $\mathcal{H}$, a UPB is a set $S$
of pure orthogonal product states spanning a proper subspace
$\mathcal{H}_s$ of $\mathcal{H}$, whose complementary subspace
$\cH^\perp_s$ contains no product states. UPB has been known to have
intimate relationship with bound entangled states, which has
positive partial transpose. For these reasons UPB are extensively
studied~\cite{dms03,AL01,CJ13,Joh13}. The generalization of UPB by
allowing non-orthogonal product state has been also shown to be
relevant~\cite{Pit03,Par04,rajaramabhat06,SLJM10}.

To generalize the idea of UPBs by using the Slater determinant as an analogue of product
states was first suggested in~\cite{sck01}, where it was suggested that
direct generalization using existing UPBs in the tensor product case
by antisymmetrying the corresponding product state could not work,
as all the known UPBs involve product states of the form $\ket{\psi}\otimes\ket{\xi}$,
with $\ket{\psi}$ and $\ket{\xi}$ being nonorthogonal.
Therefore the construction of fermionic UPBs needs significantly different
considerations that the tensor product case.

The goal of this work is to study UPBs in fermionic systems. As already mentioned, we will
use Slater determinants as product states for fermionic systems. It might be more precise
to use the name `unextendable Slater determinant bases' in this case, as similarly suggested in
~\cite{sck01} for bosonic systems, we decide to keep the name UPB to emphasize its analogue
to the original UPB. We study both the orthogonal version and the generalized version. In both cases,
analogues to the original UPB are found, but significant differences also do exist.

\section{Preliminaries}

\subsection{Unextendible Product Basis}

We consider a N-partite quantum system with Hilbert space
$\cH=\bigox^N_{i=1}\cH_i$ with $\dim\cH_i=d_i$. A {\it product
state} $\ket{\psi}\in\cH$ is a state with the form
$\ket{\psi}=\bigotimes_{i=1}^N\ket{\alpha_i}$ with
$\ket{\alpha_i}\in\cH_i$.

\bd \label{df:CES} A subspace $\mathcal{S}\subset\cH$ is completely
entangled if it does not contain any product vector. \ed

The following definition is given by \cite[Definition
1]{dms03}.
 \bd
 \label{df:UPB}
An unextendible product basis (UPB) is a set of orthogonal product
states spanning a subspace of $\mathcal{H}$, whose orthogonal
complement is completely entangled. \ed

 \bd
 \label{df:GUPB}
A generalized UPB is a set of linearly independent product states
spanning a subspace of $\mathcal{H}$, whose orthogonal complement is
completely entangled.
 \ed

Evidently a UPB must be a generalized UPB, but the converse is
false. Since $\cH$ is spanned by the computational basis states
$\{\ket{i_1,\cdots,i_N}\}$, they form the \textit{trivial} UPB. On
the other hand, the orthogonal basis spanning $\cH$ is not
necessarily equivalent to the trivial UPB. To construct such a
basis, we consider the following matrices:
 \bea
 &&
 A_1,~~~\ket{a_{i_1}},~~~ d_1 \times d_1,
 \notag\\
 &&
 A_2,~~~\ket{a_{i_1i_2}},~~~ d_2 \times d_1d_2,
 \notag\\
 &&
 \cdots,
 \notag\\
 &&
 A_{N},~~~\ket{a_{i_1i_2\cdots i_{N}}},~~~d_{N} \times (d_1d_2\cdots
 d_N).
 \eea
Here, $A_k$ is a $d_k\times(d_1d_2\cdots d_k)$ matrix with columns
$\ket{a_{i_1i_2\cdots i_k}}$, $i_j=1,\cdots,d_j$ for $j=1,\cdots,k$.
Any $A_k$ can be consecutively divided by columns into $d_1d_2\cdots
d_{k-1}$ unitary submatrices of size $d_k\times d_k$. Then one can
check that the product states
$\ket{a_{i_1},a_{i_1i_2},\cdots,a_{i_1i_2\cdots i_{N}}}$, $\forall
i_1,\cdots,i_N$ are orthonormal and span $\cH$, so they form a UPB.
By choosing $A_1,\cdots,A_N$ with pairwise different columns, we can
show that the UPB is not equivalent to the trivial UPB. Besides,
such a UPB can be real by choosing real $A_1,\cdots,A_N$.

In this paper we will use mainly the \textit{nontrivial} UPB and
generalized UPB whose cardinality is smaller than $\dim\cH$. The
following result is known for generalized UPB~\cite{rajaramabhat06}.
 \bl
 \label{le:GUPB}
Let $L_N=\sum^N_{i=1} d_i - N + 1$ and $D_N=d_1\times d_2\times
\cdots \times d_N$. Then for all integers $c \in[L_N,D_N]$, there is
a generalized UPB of cardinality $c$.
 \el

From this result, it follows that the maximum dimension of a
completely entangled subspace of $\cH$ is $D_N-L_N$~\cite{Par04}.
Since any subspace of a completely entangled subspace is still
completely entangled, for all $c\in[1,D_N-L_N]$ there is a
completely entangled subspace of dimension $c$.

Unfortunately, there is no similar general results as
Lemma~\ref{le:GUPB} for UPBs. The minimum size of UPB was first
studied by Alon and Lov\'{a}sz in 2001 \cite{AL01} and then
attracted a lot of interests \cite{Fen06,Ped02}. However, even the
bipartite case was not fully understood until very recently
\cite{CJ13}.  Let's denote the minimum size of UPB in $\cH$ by
$f_m(d_1,d_2,\cdots,d_N)$. Then for the bipartite case, we have
\begin{eqnarray*}
f_m(d_1,d_2)= \left\{\begin{array}{ll}
d_1d_2 & \textrm{if $\min\{d_1,d_2\}=2$;}\\
d_1+d_2 & \textrm{if $d_1,d_2\geq 4$ are even;}\\
d_1+d_2-1 & \textrm{otherwise.}
\end{array} \right.
\end{eqnarray*}

In the multipartite scenario, though in many cases, the minimum size of a UPB does not exceed the trivial lower bound $L_N$ by more than $1$, it is still unclear whether this is always the case.

\subsection{Fermionic Systems}

Let $V$ be a complex Hilbert space of dimension $M$ and
$\cH=\otimes^N V$ the $N$th tensor power of $V$. When there is no ambiguous, we will use $\mathbb{C}^M$ instead of $V$ for the sake of simplicity. The inner product
on $V$ extends to one on $\cH$ such that
$$
\braket{v_1,v_2,\ldots,v_N}{w_1,w_2,\ldots,w_N}=
\prod_{i=1}^N \braket{v_i}{w_i}.
$$
We denote by $\wedge^N V$ the $N$th exterior power of $V$, i.e., the
subspace of $\cH$ consisting of the antisymmetric tensors. We refer
to vectors $\ket{\psi}\in\wedge^N V$ as $N$-{\em vectors}. For any
vectors $\ket{v_i}\in V$, $i=1,\ldots,N$, we shall use the standard
algebraic notation
 \bea
 \label{ea:isomorphism}
 |v_1\rangle\wedge \cdots
 \wedge|v_N\rangle := \sum_{\sigma\in
 S_N}\mathrm{sgn}(\sigma)|v_{\sigma(1)},v_{\sigma(2)},\ldots,v_{\sigma(N)}\rangle,
 \eea
where $S_N$ is the symmetric group on $N$ letters and
$\mathrm{sgn}(\sigma)$ the sign of the permutation $\sigma$.
We say that an $N$-vector $\ket{\psi}$ is {\em decomposable} if
it can be written as
$$
\ket{\psi}=|v_1\rangle\wedge \cdots  \wedge|v_N\rangle
$$
for some $\ket{v_i}\in V$.

The inner product of two decomposable $N$-vectors
$\ket{v}=|v_1\rangle\wedge\cdots\wedge|v_N\rangle$ and
$\ket{w}=|w_1\rangle\wedge\cdots\wedge|w_N\rangle$ is equal to the
determinant of the $N\times N$ matrix $[\braket{v_i}{w_j}]$.

If $W$ is a vector subspace of $V$, then $\wedge^N W$ is a vector
subspace of $\wedge^N V$. Given a $\ket{\psi}\in\wedge^N V$, there
exists the smallest subspace $W\sue V$ such that
$\ket{\psi}\in\wedge^N W$. We shall refer to this subspace as the
{\em support} of $\ket{\psi}$ and denote it by $\sp(\ps)$. It
follows from \eqref{ea:isomorphism} that in the tensor product
space, $\sp(\ps)$ is the range of reduced density operator of any
single system. In the case when $\ket{\psi}\ne0$ is decomposable,
say $\ket{\psi}=|v_1\rangle\wedge \cdots \wedge|v_N\rangle$, then
its support is the subspace spanned by the vectors $\ket{v_i}$,
$i=1,\ldots,N$.

As already discussed, decomposable vectors are analogues of
product states in fermionic systems. The two names will
be used interchangeably.

The 2-vectors $\ket{\psi}\in\we^2 V$ are often identified with
antisymmetric matrices of order $M$ via the isomorphism which
assigns to any decomposable 2-vector
$\ket{\psi}=\ket{v}\we\ket{w}$ the matrix
$\ket{v}(\ket{w}^T)-\ket{w}(\ket{v}^T)$. Under this isomorphism,
the nonzero decomposable 2-vectors correspond to antisymmetric
matrices of rank 2.

If $A_i:V\to V$, $i=1,\ldots,N$, are linear operators, then their
tensor product $\otimes_{i=1}^N A_i$ will be identified with the
unique linear operator on $\cH$ which maps
$$
 \ket{v_1}\otimes\ket{v_2}\otimes
\cdots\otimes\ket{v_N} \to
A_1\ket{v_1}\otimes A_2\ket{v_2}\otimes
\cdots\otimes A_N\ket{v_N},
\quad \ket{v_i}\in V,\quad(i=1,2,\ldots,N).
$$
If $A_1=A_2=\cdots=A_N=A$, then we shall write $\otimes^N A$
or $A^{\otimes N}$ instead of $\otimes_{i=1}^N A_i$ and refer to
$\otimes^N A$ as the $N$th tensor power of $A$.
If $\ket{\psi}\in\wedge^N V$ then we also have
$\otimes^N A(\ket{\psi})\in\wedge^N V$. Consequently, we
can restrict the operator $\otimes^N A$ to obtain a linear
operator on $\wedge^N V$, which we denote by $\wedge^N A$ or
$A^{\wedge N}$. We refer to $\wedge^N A$ as the $N$th exterior
power of $A$. Explicitly, we have
$$
\wedge^N A (\ket{v_1}\wedge\ket{v_2}\wedge
\cdots\wedge\ket{v_N})=
(A\ket{v_1})\wedge(A\ket{v_2})\wedge
\cdots\wedge(A\ket{v_N}).
$$

The general linear group $G:=\GL(V)$ acts on $\cH$ by the
so called {\em diagonal action}:
$$
A\cdot(\ket{v_1}\otimes\ket{v_2}\otimes
\cdots\otimes\ket{v_N})=
A\ket{v_1}\otimes A\ket{v_2}\otimes
\cdots\otimes A\ket{v_N}, \quad A\in G,
\quad \ket{v_i}\in V,\quad(i=1,2,\ldots,N).
$$
In other words, $A\in G$ acts on $\cH$ as $\otimes^N A$. By
restriction, we obtain an action of $G$ on $\wedge^N V$ where $A\in
G$ acts as $\wedge^N A$.

We shall say that two $N$-vectors $\ket{\phi}$ and $\ket{\psi}$ are
{\em equivalent} if they belong to the same $G$-orbit, i.e.,
$\ket{\psi}=A\cdot\ket{\phi}$ for some $A\in G$. We shall also say
that they are {\em unitarily equivalent}, or equivalent up to an
local unitary (LU) if such $A$ can be chosen to be unitary. For
example, any decomposable $N$-vector is unitarily equivalent to a
scalar multiple of $\ket{1}\we\cdots\we\ket{N}$.

In the case $N=2$, we have the canonical form for unitary
equivalence.
 \bl
 \label{le:2ferM}
If $N=2$, then any 2-vector $\ket{\psi}$ is unitarily equivalent
to $\sum^k_{i=1}c_i \ket{2i-1}\we\ket{2i}$ for some
$c_1\ge\cdots\ge c_k>0$ and some integer $k\ge0$. Moreover, the
coefficients $c_i$ are uniquely determined by $\ket{\psi}$.
 \el
 \bpf
The first assertion follows easily from the antisymmetric version
of Takagi's theorem, see \cite[p. 217]{hj85}.
To prove the second assertion, we denote by $K$ the antisymmetric
matrix corresponding to $\ket{\psi}$. If $U$ is a unitary
operator on $V$, then the 2-vector $U\ket{\psi}$ is represented
by the antisymmetric matrix $UKU^T$. It follows that the
Hermitian matrix $KK^*$ is transformed to $UKK^*U^\dag$, and so
its eigenvalues are independent of $U$. Since these eigenvalues
are $-c_1^2,-c_1^2,\ldots,-c_k^2,-c_k^2,0,\ldots,0$, the second
assertion follows.
 \epf

From this result we have
 \bcr
 \label{cr:PS2wedgePS2}
A 2-vector $\ket{\ps}$ is decomposable if and only if
$\wetw{\ps}{\ps}=0$.
 \ecr
The corollary cannot be generalized to $n$-vectors $\ket{\ps}$ with
$n>2$. Indeed, it is easy to check that the equation
$\wetw{\ps}{\ps}=0$ is satisfied for any odd $n$.

Unless stated otherwise, the states will not be normalized for
convenience. We shall use the above properties and conventions in
this paper.

\section{\label{sec:GFUPB} Generalized fermionic UPB}

We investigate the UPB concept in fermionic systems, by treating
decomposable states such as the Slater determinants states
$\{e_{i_1,\cdots,i_N}:=\ket{i_1}\we\cdots\we\ket{i_N}\}$ as an
analogue of product states. We start from the simple case where we
do not require orthogonality of decomposable states in this section,
then move on to the orthogonal case in the next section.

We work with the $N$-fermion Hilbert space $\we^N \bC^M$. We have
the following definitions similar to Definition \ref{df:CES},
\ref{df:UPB} and \ref{df:GUPB}.

\bd \label{df:FCES} A subspace $\mathcal{S}\sue\we^N \bC^M$ is completely
entangled if it does not contain any decomposable vector. \ed

 \bd \label{df:FUPB} A ferminionic unextendible product basis (FUPB)
in $\we^N \bC^M$ is an orthogonal set of decomposable vectors, whose
orthogonal complement is completely entangled.
 \ed

\bd \label{df:gFUPB} A generalized FUPB is a set of linearly
independent decomposable vectors whose orthogonal complement is
completely entangled. \ed

Evidently an FUPB is also a generalized FUPB, but the converse is
false. Since $\we^N \bC^M$ is spanned by the Slater determinant
basis, they form the \textit{trivial} FUPB. This is also the only
FUPB up to overall factors of FUPB members when $M=N$.

In this paper we will investigate only the \textit{nontrivial} FUPB
and generalized FUPB whose cardinality is smaller than $\dim\we^N \bC^M$.
The first question we will ask is to find an analogue of
Lemma~\ref{le:GUPB}. It turns out that the question of computing the
maximum dimension of completely entangled subspaces for fermions can
be answered by algebraic geometry methods. For instance, the set of decomposable states for an
$N$-fermion system in an $M$-dimensional space is a projective
variety defined by Pl{\"u}cker's equations, of dimension $N(M-N)$. To see this,
any normalized decomposable state $|v_1\rangle\wedge \cdots  \wedge|v_N\rangle
$ is associated with a $N$-dimensional subspace spanned by $\{|v_1\rangle, \cdots ,|v_N\rangle\}$.
Notice that the change of basis will not affect the state. Thus, the set of decomposable
states in $\we^N \bC^M$ is isomprhic to the set of all $N$-dimensional subspace of $\bC^M$, which was known as the Grassmannian $Gr_N(\bC^M)$, a well-studied object in both algebraic and differential geometry.

Let $S$ be a $N$-dimensional subspace spanned by $\{|v_1\rangle, \cdots ,|v_N\rangle\}$. For $1\leq j\leq N$, let's express $|v_j\rangle$ as a sum of the computational vectors scaled $\sum\limits_{i=1}^M a_{ij}|i\rangle$. Then the coordinates of $|v_1\rangle\wedge \cdots  \wedge|v_N\rangle$ are called the Pl{\"u}cker coordinates. In fact, these are nothing but the ${M\choose N}$ minors of the matrix $(a_{ij})$. The Pl{\"u}cker coordinates satisfy some simple quadratic polynomials called the Pl{\"u}cker relations. For more details, please refer to \cite{h95}.

It is also known that the dimension of an irreducible projective variety $X\sue\cP^n$ is the
smallest integer $k$ such that there exists a linear subspace of
dimension $n-k-1$ disjoint from $X$ \cite{h95}. Therefore we have
the following proposition.
 \begin{lemma}
 \label{le:FCES}
The maximum dimension for a completely entangled subspace
$\mathcal{S}\sue \we^N \bC^M$ is
\begin{equation}
\label{eq:dim} \dim\cS={M\choose N}-N(M-N)-1.
\end{equation}
 \end{lemma}

This directly gives us the smallest possible dimension of
generalized FUPBs, namely $N(M-N)+1$.

 \begin{lemma}
 \label{cr:>=d-sPRODstate}
Let $\cF\subseteq \we^N \bC^M$ be a subspace of dimension $d\ge s=
{M\choose N}-N(M-N)-1$. Then $\cF$ contains at least $d-s$
orthogonal decomposable states.
 \end{lemma}
 \bpf
We use induction on $d$. The case $d=s$ is trivial. Let $d>s$. By
Lemma \ref{le:FCES} there is a decomposable vector
$\ket{\ph}\in\cF$. Let $\cF'$ be a hyperplane of $\cF$ such that
$\ket{\ph}\perp\cF'$. We can now apply the induction hypothesis to
$\cF'$ to conclude the proof.
 \epf

 \begin{lemma}
 \label{cr:FintersecSn}
Let $\cS_1,\ldots,\cS_n$ be arbitrary $(M-N)$-dimensional subspaces of
$\bC^M$. If $n\le N(M-N)$ then there exists an N-dimensional
subspace $\cF$ such that $\cF\cap\cS_i\ne\{0\}$ for all $i$.
 \end{lemma}
 \bpf
We construct $n$ decomposable states $\ket{\ps_i}$ in $\we^N\bC^M$,
such that the space of $\ket{\ps_i}$ is orthogonal to
$\cS_i,i=1,\cdots,n$. Since $n\le N(M-N)$, it follows from Lemma
\ref{le:FCES} that there is a decomposable state $\ket{f}$
orthogonal to $\ket{\ps_i},\forall i$. Then the space of $\ket{f}$,
named as $\cF$, properly intersects with any $\cS_i$. Since $\cF$ is
$N$-dimensional, this completes the proof.
 \epf

Our next goal is to find a generalized FUPB which has the minimal
possible cardinality $N(M-N)+1$. Following up the Corollary above,
we further claim that when $n=N(M-N)+1$, there exist subspaces
$\cS_1,\ldots,\cS_n$ such that there is no N-dimensional subspace $\cF$
such that $\cF\cap \cS_i\ne\{0\}$ for all $i$. By regarding $\cS_i^\perp$
as the support of the decomposable state $\ket{\ps_i},i=1,\cdots,n$,
these states would form a generalized FUPB. It is comparable to that in the
tensor product space introduced in Lemma \ref{le:GUPB}. What's more,
the subspace orthogonal to the space spanned by
$\ket{\ps_i},i=1,\cdots,n$ is a completely entangled subspace of maximum dimension as given in Eq.
\eqref{eq:dim}.

To prove the claim, it follows from Lemma \ref{le:2ferM} that, for
$N=2$ and $M=2$ or $3$, any pure fermionic state can be written as a
single slater determinant, hence is separable. So there is no
nontrivial generalized FUPB in $\we^2\bC^2$ or $\we^2\bC^3$. For
arbitrary $M\ge4,N\ge2$ we have

\begin{theorem}
 \label{thm:GFUPB}
The following $N(M-N)+1$ decomposable states form a generalized FUPB
in $\we^N \bC^M$.
\begin{eqnarray}
 &&(\ket{1}+t\ket{2}+t^2\ket{3}+\cdots +t^{M-1}\ket{M})
 \\
 &\wedge& (\ket{1}+(t+1)\ket{2}+(t+1)^2\ket{3}+\cdots +(t+1)^{M-1}\ket{M})\nonumber
 \\
 &\wedge& \cdots
 \\
 &\wedge&
 (\ket{1}+(t+N-1)\ket{2}+(t+N-1)^2\ket{3}+\cdots +(t+N-1)^{M-1}\ket{M})
\end{eqnarray}
where $t$ ranges from $1$ to $N(M-N)+1$.
\end{theorem}

\begin{proof}
If there is some decomposable state $(\sum\limits_{i=1}^M x_{i1}
\ket{i}) \wedge (\sum\limits_{j=1}^M x_{j2}\ket{j})\wedge \cdots
\wedge (\sum\limits_{k=1}^M x_{kN}\ket{k} )$ orthogonal to the above
$N(M-N)+1$ decomposable states, then follows from the fact that the
inner product of two decomposable $N$-vectors
$\ket{v}=|v_1\rangle\wedge\cdots\wedge|v_N\rangle$ and
$\ket{w}=|w_1\rangle\wedge\cdots\wedge|w_N\rangle$ is equal to the
determinant of the $N\times N$ matrix $[\braket{v_i}{w_j}]$, we will
have the following equations.

\begin{eqnarray}
\det \left(
\begin{array}{cccc}
\sum\limits_{j=1}^M x_{j1}t^{j-1}  & \sum\limits_{j=1}^M x_{j2}t^{j-1}  & \cdots  & \sum\limits_{j=1}^M x_{jN}t^{j-1}\\
\sum\limits_{j=1}^M x_{j1}(t+1)^{j-1}  & \sum\limits_{j=1}^M x_{j2}(t+1)^{j-1} & \cdots  & \sum\limits_{j=1}^M x_{jN}(t+1)^{j-1}\\
\vdots  &  \vdots  & \ddots  & \vdots\\
\sum\limits_{j=1}^M x_{j1}(t+N-1)^{j-1}  & \sum\limits_{j=1}^M
x_{j2}(t+N-1)^{j-1}  & \cdots  & \sum\limits_{j=1}^M
x_{jN}(t+N-1)^{j-1}
\end{array}
\right)=0
\end{eqnarray}
where $t$ ranges from $1$ to $N(M-N)+1$.

Or equivalently,
\begin{eqnarray}
\det \Bigg ( \left(
\begin{array}{cccc}
1& t&\cdots & t^{M-1}\\
1& t+1&\cdots & (t+1)^{M-1}\\
\vdots & \vdots & \ddots & \vdots \\
1& t+N-1&\cdots & (t+N-1)^{M-1}
\end{array}\right)
\cdot\left(
\begin{array}{cccc}
x_{11}& x_{12}&\cdots & x_{1N}\\
x_{21}& x_{22}&\cdots & x_{2N}\\
\vdots & \vdots & \ddots & \vdots \\
x_{M1}& x_{M2}&\cdots & x_{MN}
\end{array}\right)
\Bigg )=0
\end{eqnarray}
where $t$ ranges from $1$ to $N(M-N)+1$.

To continue our proof, we need the Binet-Cauchy formula which states that, for any $m\times n$ matrix $A$ and $n\times m$ matrix $B$,
\begin{eqnarray*}
\det(AB)=\sum\limits_I \det(A_I)\det(B_I)
\end{eqnarray*}
where $I$ ranges over all subsets of $\{1,2,\cdots,n\}$ with size $m$, $A_I$ is the submatrix of $A$ formed by the columns indexed by $I$ and $B_I$ is the submatrix of $B$ formed by the rows indexed by $I$.

By applying the Binet-Cauchy formula introduced above, we have
\begin{eqnarray}
g(t)&=&\sum\limits_{1\leq p_1<p_2<\cdots <p_N\leq M} \det \left(
\begin{array}{cccc}
t^{p_1-1}& t^{p_2-1}&\cdots & t^{p_N-1}\\
(t+1)^{p_1-1}& (t+1)^{p_2-1}&\cdots & (t+1)^{p_N-1}\\
\vdots & \vdots & \ddots & \vdots \\
(t+N-1)^{p_1-1}& (t+N-1)^{p_2-1}&\cdots & (t+N-1)^{p_N-1}\nonumber\\
\end{array}\right)\\
&\cdot& \det \left(
\begin{array}{cccc}
x_{p_11}& x_{p_12}&\cdots & x_{p_1N}\\
x_{p_21}& x_{p_22}&\cdots & x_{p_2N}\\
\vdots & \vdots & \ddots & \vdots \\
x_{p_N1}& x_{p_N2}&\cdots & x_{p_NN}
\end{array}\right)
=0
\end{eqnarray}
for any integer $1\leq t\leq N(M-N)+1$.

For any $1\leq p_1<p_2<\cdots <p_N\leq M$, let's define
\begin{eqnarray}
f_{p_1,\cdots, p_N}(y_1,y_2,\cdots, y_N)&=&\det \left(
\begin{array}{cccc}
y_1^{p_1-1}& y_1^{p_2-1}&\cdots & y_1^{p_N-1}\\
y_2^{p_1-1}& y_2^{p_2-1}&\cdots & y_2^{p_N-1}\\
\vdots & \vdots & \ddots & \vdots \\
y_N^{p_1-1}& y_N^{p_2-1}&\cdots & y_N^{p_N-1}\nonumber\\
\end{array}\right)
\end{eqnarray}
which is a multivariable polynomial of degree at most
$\sum\limits_{i=1}^N p_i-N$.  In fact, $f_{p_1,\cdots,p_N}$ is a
generalization of Vandermonde determinant. Since
$f_{p_1,\cdots,p_N}(y_1,y_2,\cdots, y_N)$ vanishes on equating any
$y_i$ and $y_j(i\neq j)$, we know $f_{p_1,\cdots,
p_N}(y_1,y_2,\cdots,y_N)$ is divisible by the Vandermonde
determinant  $V_N(y_1,y_2, \cdots, y_N)=\det \left(
\begin{array}{cccc}
1& y_1&\cdots & y_1^{N-1}\\
1& y_2&\cdots & y_2^{N-1}\\
\vdots & \vdots & \ddots & \vdots \\
1& y_N&\cdots & y_N^{N-1}\nonumber\\
\end{array}\right)=\prod\limits_{1\leq i<j\leq N} (y_j-y_i)$ in the polynomial ring $\mathbb{Z}(y_1,y_2,\cdots, y_N)$.

We are mainly interested in the quotient polynomial of
$f_{p_1,\cdots,p_N}(y_1,\cdots,y_N)$ by the Vandermonde determinant
$V_N(y_1,y_2, \cdots, y_N)$. Its degree is at most
\begin{eqnarray}
\deg(f_{p_1,\cdots,p_N})-\deg (V_N)\leq \sum\limits_{i=1}^N p_i-N
-{N\choose 2}\leq \sum\limits_{i=1}^N (M-i+1)-N -{N\choose
2}=N(M-N).
\end{eqnarray}

Immediately, we have
\begin{eqnarray}
g(t)&=&\sum\limits_{1\leq p_1<p_2<\cdots <p_N\leq M} \det \left(
\begin{array}{cccc}
x_{p_11}& x_{p_12}&\cdots & x_{p_1N}\\
x_{p_21}& x_{p_22}&\cdots & x_{p_2N}\\
\vdots & \vdots & \ddots & \vdots \\
x_{p_N1}& x_{p_N2}&\cdots & x_{p_NN}
\end{array}\right) f_{p_1,\cdots,p_N}(t,\cdots, t+N-1)\nonumber
\end{eqnarray}
which implies
\begin{eqnarray}
\deg g(t)&\leq& \max\limits_{1\leq p_1<p_2<\cdots < p_\leq M}  \deg f_{p_1,\cdots,p_N}(t,\cdots, t+N-1)\\
&= & \max\limits_{1\leq p_1<p_2<\cdots < p_\leq M}  \deg f_{p_1,\cdots,p_N}/V_N (t,\cdots, t+N-1)\\
&\leq & N(M-N).
\end{eqnarray}

Recall that $g(t)=0$ for any $t=1,2,\cdots, N(M-N)+1$,  hence $g(t)$
is a zero polynomial.

Let's denote $\det \left(
\begin{array}{cccc}
x_{p_11}& x_{p_12}&\cdots & x_{p_1N}\\
x_{p_21}& x_{p_22}&\cdots & x_{p_2N}\\
\vdots & \vdots & \ddots & \vdots \\
x_{p_N1}& x_{p_N2}&\cdots & x_{p_NN}
\end{array}\right)$ by $P_{p_1,p_2,\cdots, p_N}$, then
\begin{eqnarray}
(P_{1,2,\cdots,N}, P_{1,2,\cdots, N-1,N+1}, \cdots, P_{M-N+1,\cdots,
M})
\end{eqnarray}
is the Pl\"ucker coordinate of $N$-dimensional subspace spanned by
\begin{eqnarray}
\left(
\begin{array}{c}
x_{11}\\
x_{21}\\
\vdots\\
x_{M1}
\end{array}\right),\left(
\begin{array}{c}
x_{12}\\
x_{22}\\
\vdots\\
x_{M2}
\end{array}\right),\cdots, \left(
\begin{array}{c}
x_{1N}\\
x_{2N}\\
\vdots\\
x_{MN}
\end{array}\right).
\end{eqnarray}

We also denote $\det \left(
\begin{array}{cccc}
t^{p_1-1}& t^{p_2-1}&\cdots & t^{p_N-1}\\
(t+1)^{p_1-1}& (t+1)^{p_2-1}&\cdots & (t+1)^{p_N-1}\\
\vdots & \vdots & \ddots & \vdots \\
(t+N-1)^{p_1-1}& (t+N-1)^{p_2-1}&\cdots & (t+N-1)^{p_N-1}
\end{array}\right)$ by $\Delta_{p_1,p_2,\cdots, p_N}(t)$.

To prove our theorem, we need to show that there is no nonzero
vector
\begin{eqnarray}
(P_{1,2,\cdots,N}, P_{1,2,\cdots, N-1,N+1}, \cdots, P_{M-N+1,\cdots,
M})\in \mathbf{C}^{{M\choose N}}
\end{eqnarray}
satisfying the following Pl\"ucker relations:
\begin{eqnarray}
\sum\limits_{t=1}^{N+1} (-1)^{t}
P_{j_1,j_2,\cdots,j_{N-1},j_t^{\prime}}P_{j_1^{\prime},\cdots,
j_{t-1}^{\prime},j_{t+1}^{\prime},\cdots, j_{N+1}^{\prime}}=0
\end{eqnarray}
where $\{j_t\}_{t=1}^{N-1}$ and $\{j_t^{\prime}\}_{t=1}^{N+1}$ is
any sequence in $\{1,2,\cdots, N\}$ and
\begin{eqnarray}
g(t)=\sum\limits_{1\leq p_1<p_2<\cdots <p_N\leq M}
P_{p_1,p_2,\cdots, p_N}\Delta_{p_1,p_2,\cdots, p_N}(t)=0
\end{eqnarray}
for any $t$.

We have shown that
\begin{eqnarray}
\deg \Delta_{p_1,p_2,\cdots, p_N}(t)\leq \sum\limits_{i=1}^N p_i-N
-{N\choose 2}.
\end{eqnarray}
In fact, the equality holds. Moreover, the quotient polynomial of
$f_{p_1,\cdots,p_N}(y_1,\cdots,y_N)$ by the Vandermonde determinant
$V_N(y_1,y_2, \cdots, y_N)$ contains only non-negative terms.

Therefore, monomial $t^{N(M-N)}$ appears and only appears in
$\Delta_{M-N+1,M-N+2,\cdots, M}(t)$. Therefore $P_{M-N+1,\cdots,
M}=0$.

Then, $t^{N(M-N)-1}$ appears in $\Delta_{M-N,M-N+2,\cdots, M}(t)$
only. We have $P_{M-N, M-N+2,\cdots, M}=0$.

Assume we have already proved $P_{p_1,p_2,\cdots, p_N}=0$ for any
$\sum\limits_{i=1}^N p_i\geq K+1$, we will show that
$P_{p_1,p_2,\cdots, p_N}=0$ for any $\sum\limits_{i=1}^N p_i=K$.

If there are two nonzero $P_{p_1,p_2,\cdots, p_N}$ and
$P_{p_1^{\prime},p_2^{\prime},\cdots,
p_N^{\prime}}(\sum\limits_{i=1}^N p_i=\sum\limits_{i=1}^N
p_i^{\prime}=K)$. Without loss of generality, let
$p_N^{\prime}>p_N$. The Pl\"ucker relation:
\begin{eqnarray}
\sum\limits_{t=1}^{N+1} (-1)^{t}
P_{j_1,j_2,\cdots,j_{N-1},j_t^{\prime}}P_{j_1^{\prime},\cdots,
j_{t-1}^{\prime},j_{t+1}^{\prime},\cdots, j_{N+1}^{\prime}}=0
\end{eqnarray}
where $(j_1,j_2,\cdots,j_{N-1})=(p_1^{\prime},p_2^{\prime},\cdots,
p_{N-1}^{\prime})$ and $(j_1^{\prime}, j_2^{\prime}, \cdots,
j_N^{\prime}, j_{N+1}^{\prime})=(p_1,p_2,\cdots, p_N,
p_{N}^{\prime})$ will lead to $P_{p_1,p_2,\cdots,
p_N}P_{p_1^{\prime},p_2^{\prime},\cdots, p_N^{\prime}}=0$. This
implies that there is at most one nonzero $P_{p_1,p_2,\cdots, p_N}$
for all $\sum\limits_{i=1}^N p_i=K$.

If there is exactly one nonzero $P_{p_1,p_2,\cdots, p_N}$ for all
$\sum\limits_{i=1}^N p_i=K$, then $t^{K-N-{N\choose 2}}$ is the
leading monomial of $g(t)$ and its coefficient is nonzero. This
contradicts $g(t)=0$.

Therefore, $P_{p_1,p_2,\cdots, p_N}=0$ for any $\sum\limits_{i=1}^N
p_i=K$.

By repeating this procedure, we will have $P_{p_1,p_2,\cdots,
p_N}=0$ for any $1\leq p_1<p_2<\cdots < p_N\leq M$. This completes
our proof.
 \epf

Below we extend Lemma \ref{le:FCES} to the space of spanned by
decomposable vectors.

 \bpp
 \label{pp:smallest}
(i) There exists a subspace $L\su\we^N \bC^M$ of codimension $3$
which is not spanned by decomposable $N$-vectors.

(ii) When $N=2$, every subspace $L\sue\we^N \bC^M$ of codimension
$\le2$ is spanned by decomposable bivectors.
 \epp
 \bpf
(i) Let $\phi=e_{5,6,\ldots,N+2}$. Let $L_0$ be the subspace
of codimension $4$ spanned by all basic $N$-vectors
$e_{i_1,\ldots,i_N}$, $1\le i_1<\cdots<i_N\le M$, except
$e_{ij}\we\phi$ with $(i,j)=(1,2),(2,3),(2,4),(3,4)$.
Denote by $L$ the subspace spanned by $L_0$ and
$\psi:=(e_{12}+e_{34})\we\phi$.
It is easy to verify that for any $\a\in L_0$ the
bivector $\braket{\psi+\a}{\phi}$ is indecomposable. This implies
that $\psi+\a$ is also indecomposable. Hence, (i) is proved.

(ii) We show by induction that any subspace $H\su\we^2\bC^M$ of
dimension ${M\choose2}-2$ is spanned by 2-decomposable vectors. This
is sufficient to verify the assertion. Note that
$(A^{-1}\cdot\ket{\ps})^\perp= A^\dg\cdot(\ket{\ps}^\perp)$ for
$A\in \GL$. We can perform on $H$ any diagonal action $A\in\GL$
because $H$ is spanned by decomposable 2-vectors if and only so is
$A\cdot H$. We will frequently use this hypothesis in the proof.

The assertion is evidently true for $M=3$. Since $H$ has codimension
two, by the hypothesis we may assume that there is a state
$\ket{\a}\in \we^2\bC^{M-1}\cap H^\perp$. Applying the hypothesis
with $A=A_1\op1$ of a suitable $A_1$ on $\bC^{M-1}$, we obtain that
$H$ is spanned by the linearly independent 2-vectors
 \bea
 &&
 \label{eq:K-2}
 \ket{\ps_1}, \cdots, \ket{\ps_{K-2}},
 \\
 &&
 \label{eq:K-1}
 \ket{\ps_{K-1}}+c_1 e_{1,M},
 \cdots,
 \ket{\ps_{K+M-3}}+c_{M-1} e_{M-1,M},
 \eea
where $K={M-1 \choose 2}$ and the $\ket{\ps_i}\in\we^2\bC^{M-1}$,
$i=1,\cdots,K+M-3$. Since $\ket{\a}\in \we^2\bC^{M-1}\cap H^\perp$,
the space spanned by the $\ket{\ps_i}$ has dimension at most $K-1$.
We can remove the $\ket{\ps_i}$ for $i>K-1$ by replacing the vectors
in \eqref{eq:K-1} by a suitable linear combination of the vectors in
\eqref{eq:K-2} and \eqref{eq:K-1}. Then using the hypothesis with
$A=A_2\op1$ of a suitable $A_2$ on $\bC^{M-1}$, we obtain that $H$
is spanned by the linearly independent 2-vectors
 \bea
 &&
 \label{eq:K-2}
 \ket{\ps_1'}, \cdots, \ket{\ps_{K-2}'},
 \\
 &&
 \label{eq:K-1}
 \ket{\ps_{K-1}'}+c e_{1,M},
 \\
 &&
 \label{eq:K}
 e_{2,M},
 \cdots,
 e_{M-1,M},
 \eea
where the $\ket{\ps_i'}\in\we^2\bC^{M-1}$ and $c=0$ or $1$. Using
the induction hypothesis and replacing $\ket{\ps_{K-1}'}$ by a
suitable linear combination of it and \eqref{eq:K-2}, we may assume
that all $\ket{\ps_i'}$ are decomposable 2-vectors. To prove the
assertion, it is sufficient to show that a suitable linear
combination of $\ket{\ph}=\ket{\ps_{K-1}'}+c e_{1,M}$ and other
vectors in \eqref{eq:K-2} and \eqref{eq:K} becomes a decomposable
2-vector.

Suppose $\ket{\ph}$ is not decomposable, otherwise the assertion
follows. So $c=1$. By using the hypothesis with $A=1\op A_3\op1$ of
a suitable $A_3$, we may assume $\ket{\ps_{K-1}'}=x e_{13} +
e_{23}$, and the space spanned by \eqref{eq:K} is still spanned by
them. The space spanned by \eqref{eq:K-2} contains a nonzero state
$w e_{12}+y e_{13}+z e_{23}$ with complex numbers $w,y,z$. By adding
a suitable multiple of this state, $e_{2M}$ and $e_{3M}$ to
$\ket{\ph}$, we obtain a decomposable state. This completes the
proof.
 \epf

It follows from this proposition that any hyperplane of $\we^2\bC^M$
orthogonal to an entangled state is spanned by a generalized FUPB.
It also generalizes Lemma~\ref{le:FCES} which states that any
hyperplane of $\we^2\bC^4$ is spanned by a generalized FUPB. We
concretely construct an example.
 \bex
 \label{ex:gfupbNOTfupb}
Let $N=2$ and $\ket{\ps}=\sum^{k}_{i=1}e_{2i-1,2i}$, $1\le k\le
M/2$. Then the hyperplane $H=\ket{\ps}^\perp$ is spanned by
decomposable 2-vectors. Indeed, $H$ is spanned by
 \bea
 &&
 \wetw{i}{j},
 ~~
 i<j,
 ~~
 (i,j)\ne (1,2), \cdots, (2k-1,2k),
 \\
 &&
\bigg(\sum^k_{l=1} \o^{j(l-1)} \ket{2l-1} \bigg)
 \we
\bigg(\sum^k_{l=1} \ket{2l} \bigg),
 \\
 &&
 j=1,\cdots,k-1,
 \eea
where $\o=e^{\frac{2\p i}{k}}$.
 \qed
 \eex

By following the above argument, one can similarly construct the
decomposable vectors spanning any hyperplane of $\we^2\bC^M$ for odd
$M$. In spite of Proposition \ref{pp:smallest}, we do not know
whether any subspace in $\we^N \bC^M$ of dimension ${M\choose N}-2$
with $N>2$ is spanned by decomposable vectors. If we could construct
a 18-dimensional subspace for $N=3,M=6$ not spanned by decomposable
vectors, then we could construct a subspace in $\we^N \bC^M$ of
dimension ${M\choose N}-2$ which is also not spanned by decomposable
vectors, by following the argument in Proposition \ref{pp:smallest}
(i).

%On the other hand,

% \bqu
% Does Lemma \ref{le:hyperplane} work if the word ``GFUPB" in the claim
% is replaced by ``FUPB" ? If a space is spanned by a GFUPB, can this
% space spanned by an FUPB?
% \equ

%We are in a position to show the main result of this section.
% \bt
% \label{thm:GFUPB}
%GFUPB exists in the subspace $S\su\we^N\bC^M$ of any dimension in
%$N(M-N)+1\leq d\leq {M\choose N}$.
% \et
% \bpf
%Let
%
%This completes the proof.
% \epf

\section{\label{sec:fupb24} Fermionic UPB in $\we^2\bC^4$}

In last section we have constructed a set of generalized FUPB of the
minimal cardinality $N(M-N)+1$ in $\we^N V$. It follows from Lemma
\ref{le:FCES} that the least possible cardinality of FUPB in $\we^N V$
is also $N(M-N)+1$. However we do not have a constructive example of
such FUPB yet. By definition, the vectors in FUPB are orthogonal and
decomposable, so it is more difficult to characterize the FUPB than
the generalized FUPB. In this section we investigate the FUPB in
$\we^2\bC^4$, whose cardinality is at least five by Lemma
\ref{le:FCES}. It turns out that $\we^2\bC^4$ is the simplest
bipartite subspace in which a nontrivial FUPB exists. We will also
show that real nontrivial FUPB in $\we^2\bC^4$ does not exist.

For this purpose, we propose the following preliminary lemma. We
will use the simple fact that the FUPB is still an FUPB if either we
perform an LU $\ox^N U$ on the global system, or we linearly combine
the vectors in FUPB so that the resulting vectors are still
decomposable and orthogonal.
 \bl
 \label{le:4x4fupbPRE}
Suppose the subspace $H\sue\we^2\bC^M$.
 \\
 (a) If $M=2,3$, then $H$ contains only decomposable vectors. So $H$ is spanned by orthogonal
 decomposable states and there is no nontrivial generalized FUPB in $\we^2\bC^2$ or $\we^2\bC^3$.
 \\

Let $M=4$ and $H$ be spanned by five orthogonal decomposable vectors
$\ket{\ps_i},i=1,\cdots,5$. Then the set $S$ of these vectors is not
an FUPB when one of the following conditions is satisfied:
 \\
 (b) $\ket{\ps_i}=\ket{a}\we\ket{b_i},i=1,2,3$;
 \\
 (c) $\ket{\ps_1}=\ket{b_1}\we\ket{b_2}$, $\ket{\ps_2}=\ket{b_1}\we\ket{b_3}$, and
 $\ket{\ps_3}=\ket{b_2}\we\ket{b_3}$;
 \\
 (d) $\ket{\ps_1}=\ket{b_1}\we\ket{b_2}$ and
 $\ket{\ps_2}=\ket{b_3}\we\ket{b_4}$, where the $\ket{b_i},i=1,2,3,4$
 are orthogonal;
 \\
 (e) $\ket{\ps_i}=\ket{a}\we\ket{b_i},i=1,2$.
 \el
 \bpf
(a) is clear.

(b) By performing a suitable LU on $S$, we can assume
$\ket{\ps_i}=\wetw{i}{4},i=1,2,3$ and they span $K\su H$. So
$\ket{\ps_4}$ and $\ket{\ps_5}$ belong to the orthogonal complement
$K^\perp =\we^2 \bC^3$. It follows from assertion (a) that there is
a decomposable state $\ket{\ps_6}\in K^\perp$ orthogonal to
$\ket{\ps_4}$ and $\ket{\ps_5}$. Thus $\ket{\ps_6}\perp H$, and $S$
is not an FUPB.

(c) Using an LU we can assume
$\ket{\ps_1}=\ket{1}\we(x\ket{1}+\ket{2})$,
$\ket{\ps_2}=\ket{1}\we(y\ket{1}+\ket{3})$, and
$\ket{\ps_3}=(x\ket{1}+\ket{2})\we(y\ket{1}+\ket{3})$. Since
$\ket{\ps_1}$ and $\ket{\ps_2},\ket{\ps_3}$ are orthogonal, we have
$y=x=0$. So the sum of the supports of
$\ket{\ps_1},\ket{\ps_2},\ket{\ps_3}$ is $\bC^3$. Since
$\ket{\ps_4},\ket{\ps_5}\perp\we^2\bC^3$, we have
$\ket{4}\in\sp(\ps_4)\cap\sp(\ps_5)$. So there is a decomposable
state $\ket{\ps_6}\in H$, and $S$ is not an FUPB.

(d) Using an LU we can assume $\ket{b_i}=\ket{i},i=1,\cdots,4$.
Since $\ket{\ps_1}$ and $\ket{\ps_2}$ are orthogonal to
$\ket{\ps_i},i=3,4,5$, we have
$\ket{\ps_i}=\wetw{c_i}{f_i}=(w_i\ket{1}+x_i\ket{2})\we(y_i\ket{3}+z_i\ket{4})$.
These states are also orthogonal, so either two of the states
$\ket{c_i},i=3,4,5$ or two of the states $\ket{f_i},i=3,4,5$ are
parallel. Without loss of generality, we may assume
$\ket{c_3}=\ket{c_4}$. Since $\ket{\ps_5}$ is orthogonal to
$\ket{\ps_3}$ and $\ket{\ps_4}$, we have $\ket{c_3}\perp\ket{c_5}$.
By choosing $\ket{f_6}\in\lin\{\ket{3},\ket{4}\}$ and
$\ket{f_6}\perp\ket{f_5}$, we have $\wetw{c_5}{f_6}\perp H$. Thus
$S$ is not an FUPB.

(e) Suppose $S$ is an FUPB. Using an LU we can assume
$\ket{\ps_1}=\wetw{1}{2}$ and $\ket{\ps_2}=\wetw{1}{3}$. Since
$\ket{\ps_1}$ is orthogonal to $\ket{\ps_i},i=3,4,5$, we have
 \bea
 \label{ea:PSI345}
 \ket{\ps_3}
 &=&
 \bigg( \sum^4_{i=1}c_i\ket{i} \bigg)
 \we
 \bigg(c_5\ket{3} + c_6\ket{4} \bigg),
 \notag\\
 \ket{\ps_4}
 &=&
 \bigg( \sum^4_{i=1}d_i\ket{i} \bigg)
 \we
 \bigg(d_5\ket{3} + d_6\ket{4} \bigg),
 \notag\\
 \ket{\ps_5}
 &=&
 \bigg( \sum^4_{i=1}e_i\ket{i} \bigg)
 \we
 \bigg(e_5\ket{3} + e_6\ket{4} \bigg).
 \eea
Since $\ket{\ps_2}$ is orthogonal to $\ket{\ps_i},i=3,4,5$, we have
$c_1c_5=d_1d_5=e_1e_5=0$. If $c_1=d_1=e_1=0$, then $\wetw{1}{4}\perp
H$ and it gives us a contradiction with the hypothesis that $S$ is
an FUPB. So the case $c_1=d_1=e_1=0$ is excluded. Next the case
$c_5=d_5=e_5=0$ is excluded by claim (b). Up to the permutation of
subscripts and overall factors, it suffices to consider only two
cases: (e1) $c_1=d_5=e_5=0,c_5=1$, and (e2)
$c_1=d_1=e_5=0,c_5=d_5=1$.

In case (e1), we may assume $c_3=d_4=e_4=0$. Claim (d) implies that
$c_2=1$ up to an overall factor. Note that $c_4=c_6=0$ is excluded
by claim (c). Then the orthogonality of $\ket{\ps_i},i=3,4,5$
implies $d_2e_3=d_3e_2$. So a suitable linear combination of
$\ket{\ps_4}$ and $\ket{\ps_5}$ produces two orthogonal decomposable
states $\wetw{1}{4}$ and $\ket{\ps_5'}$. They and
$\ket{\ps_i},i=1,2,3$ still form an FUPB. It gives us a
contradiction with claim (b), so case (e1) is excluded.

In case (e2), we may assume $c_3=d_3=e_4=0$. Claim (d) implies that
$c_2=d_2=1$ up to an overall factor. Note that $e_2=e_3=0$ is
excluded by claim (b). Then the orthogonality of
$\ket{\ps_i},i=3,4,5$ implies $c_4d_6=d_4c_6$. So a suitable linear
combination of $\ket{\ps_3}$ and $\ket{\ps_4}$ produces two
orthogonal decomposable states $\wetw{2}{3}$ and $\ket{\ps_5'}$.
They and $\ket{\ps_i},i=1,2,5$ still form an FUPB. It gives us a
contradiction with claim (c), so case (e2) is also excluded.

So the hypothesis that $S$ is an FUPB is wrong. This completes the
proof.
 \epf

Evidently, claims (b) and (c) are the corollaries of (e). It is
sufficient to use claims (a), (d), (e) in the paper.

We say a set of states is real if there is an LU which converts the
set into another set whose states have real components. For example,
the trivial FUPB is real. One may expect to construct nontrivial
real FUPBs in $\we^2\bC^4$. However we have
 \bpp
 \label{pp:4x4real}
The real FUPB of five orthogonal decomposable states in $\we^2\bC^4$
does not exist.
 \epp
 \bpf
Suppose the set $S$ is a real FUPB in $\we^2\bC^4$. So we may assume
that $S$ consists of five orthogonal decomposable states
$\ket{\ps_i}=\wetw{a_i}{b_i},i=1,\cdots,5$ where the states
$\ket{a_i},\ket{b_i}$ are real. The real linear combination of
$\ket{a_i},\ket{b_j}$ is also a real vector. We will perform real
local orthogonal matrices $U\ox U$ on the states
$\ket{\ps_i},i=1,\cdots,5$, so that the resulting states still form
a real FUPB. They will still be called $\ket{\ps_i},i=1,\cdots,5$
for convenience. We shall use these conventions in the proof.

Since the $\ket{\ps_i}$ are orthogonal, by performing a suitable
real local orthogonal matrix we can assume
$\ket{\ps_1}=\wetw{1}{2}$,
$\ket{\ps_2}=(a\ket{2}+b\ket{3})\we\ket{4}$, and
$\ket{\ps_3},\ket{\ps_4},\ket{\ps_5}$ with expressions in Eqs.
\eqref{ea:PSI345}. Then Lemma \ref{le:4x4fupbPRE} (d) implies
$a\ne0$, and (e) implies that $b,c_5,d_5,e_5\ne0$. So we can assume
$a=c_5=d_5=e_5=1$. If $c_6=0$, then $\braket{\ps_2}{\ps_3}=0$
implies $c_4=0$. Since the vector
$(c_1,c_2,0,0)^T\in\sp(\ps_1)\cap\sp(\ps_3)$, it gives a
contradiction with Lemma \ref{le:4x4fupbPRE} (e). So $c_6\ne0$. The
similar argument implies $d_6,e_6\ne0$. Hence we may assume
$c_4=d_4=e_4=0$. Next, Lemma \ref{le:4x4fupbPRE} (e) implies that
$c_3d_3e_3\ne0$. Using the equation $\braket{\ps_2}{\ps_i}=0$,
$i=3,4,5$, we can assume $(c_2,c_3)=(d_2,d_3)=(e_2,e_3)=(-b,1)$. Now
$S$ consists of five orthogonal states
 \bea
 \label{eq:4by4FUPB}
 \ket{\ps_1} &=&
 \wetw{1}{2},
 \notag\\
 \ket{\ps_2} &=&
 (\ket{2}+b\ket{3})\we\ket{4},
 \notag\\
 \ket{\ps_3} &=&
 (c_1\ket{1}-b\ket{2}+\ket{3})\we(\ket{3}+c_6\ket{4}),
 \notag\\
 \ket{\ps_4} &=&
 (d_1\ket{1}-b\ket{2}+\ket{3})\we(\ket{3}+d_6\ket{4}),
 \notag\\
 \ket{\ps_5} &=&
 (e_1\ket{1}-b\ket{2}+\ket{3})\we(\ket{3}+e_6\ket{4}).
 \eea
If $c_1=0$, then $\sp(\ps_2)\cap\sp(\ps_3)\ne\{0\}$. It is a
contradiction with Lemma \ref{le:4x4fupbPRE} (e), so $c_1\ne0$.
Similar arguments show $d_1,e_1\ne0$. The equations
$\braket{\ps_3}{\ps_4}=\braket{\ps_3}{\ps_5}=\braket{\ps_4}{\ps_5}=0$
imply
 \bea
 \label{ea:c1d1e1}
 c_1 d_1
 &=&
 \frac{1}{1+c_6d_6} - b^2 - 1,
 \\
 c_1 e_1
 &=&
 \frac{1}{1+c_6e_6} - b^2 - 1,
 \\
 \label{ea:d1e1}
 d_1 e_1
 &=&
 \frac{1}{1+d_6e_6} - b^2 - 1.
 \eea
Since $b,c_1,d_1,e_1,c_6,d_6,e_6\ne0$ are real, let
$f_1=\frac{b^4+b^2}{c_1d_1e_1}$ and
$f_6=\frac{b^2}{c_6d_6e_6(1+b^2)}$. One can directly check that the
decomposable state
$(f_1\ket{1}-b\ket{2}+\ket{3})\we(\ket{3}+f_6\ket{4})$ is orthogonal
to the states $\ket{\ps_1},\cdots,\ket{\ps_5}$. It gives us a
contradiction with the assumption that $S$ is an FUPB.

So there is no nontrivial real FUPB in $\we^2\bC^4$. This completes
the proof.
 \epf

In spite of the above findings, we show that a complex FUPB in
$\we^2\bC^4$ indeed exists.

 \bex
 \label{ex:4x4FUPB}
 {\rm
We consider the following set $S$ of five decomposable states
in $\we^2\bC^4$:
 \bea
 \ket{\ps_1} &=& \wetw{1}{2},
 \\
 \ket{\ps_2} &=& (\ket{2}-b\ket{3})\we\ket{4},
 \\
 \ket{\ps_3}
 &=&
 \bigg( \ket{1}+b\ket{2}+\ket{3} \bigg)\we (\ket{3}+\ket{4}),
 \\
 \ket{\ps_4}
 &=&
 \bigg( (-\frac{c}{1+c} - b^2) \ket{1}+b\ket{2}+\ket{3} \bigg)\we (\ket{3}+c\ket{4}),
 \\
 \ket{\ps_5}
 &=&
 \bigg( (-\frac{d}{1+d} - b^2) \ket{1}+b\ket{2}+\ket{3} \bigg)\we (\ket{3}+d\ket{4}),
 \eea
where $b>0$, and $c, d$ are nonzero and complex. We require
$\ket{\ps_4}$ and $\ket{\ps_5}$ are orthogonal, i.e.,
$\braket{\ps_4}{\ps_5}=0$. This is a quadratic equation of $c$. We
claim that the equation has double root by choosing suitable nonzero
scalars $b,d$. For example, we may approximately take $b=2,
d\approx1.13631 - 0.197693i$. The corresponding double root is
$c\approx-0.829747 + 0.0716405 i$.

Suppose there is a nonzero decomposable vector $\ket{\ph}$
orthogonal to the five states in the set $S$ with the above $b,c,d$.
So the state has the form
$\ket{\ph}=(w\ket{1}+c_2\ket{2}+x\ket{3}+c_4\ket{4})\we(y\ket{3}+z\ket{4})$.
The equations $\braket{\ps_i}{\ph}=0,i=2,3,4$ imply that $z\ne0$. So
we can assume $c_4=0$. Then the equation $\braket{\ps_2}{\ph}=0$
implies $c_2=2x$. Now the state becomes
 \bea
 \ket{\ph}
 =
 \left(
   \begin{array}{c}
     w \\
     2x \\
     x \\
     0 \\
   \end{array}
 \right)
 \we
 \left(
   \begin{array}{c}
     0 \\
     0 \\
     y \\
     z \\
   \end{array}
 \right).
 \eea
Since $\ket{\ph}$ is orthogonal to $\ket{\ps_i},i=3,4$, we obtain
$y\ne0$. So we can assume $y=1$ up to an overall factor. By the same
reason we have $x=1$. The equation $\braket{\ph}{\ps_3}=0$ implies
that $w=-\frac{z}{1+z} - 4$. Therefore, the two states $\ket{\ph}$
and $\ket{\ps_4}$ are the same when $z=c$. Recall that
$\braket{\ps_4}{\ps_5}=0$ has a unique solution $c$. So the equation
$\braket{\ph}{\ps_5}=0$ implies that $\ket{\ph}$ and $\ket{\ps_4}$
are parallel. It is a contradiction with the assumption that they
are orthogonal.

So there is no decomposable vector orthogonal to the states in $S$.
On the other hand, it is easy to check that the five decomposable
states $\ket{\ps_i},i=1,\cdots,5$ are orthogonal. So the set $S$ is
an FUPB.
 \qed }
 \eex

Using the same method of requiring double roots in the equation
$\braket{\ps_4}{\ps_5}=0$, one may construct many other examples of
FUPB in $\we^2\bC^4$. By modifying the expression of $\ket{\ps_3}$
with a more general form, it is possible to characterize all in
$\we^2\bC^4$ FUPB up to LU. Finally even if the equation
$\braket{\ps_4}{\ps_5}=0$ does not have double roots and has close
enough different roots, we may still construct an FUPB, by following
a similar argument to Example \ref{ex:4x4FUPB}.

% \bqu
%Can we obtain a canonical form of FUPB in $\we^2\bC^4$ based on Example
%\ref{ex:4x4FUPB} ?
%\equ

\section{\label{sec:fupbhigh} Fermionic UPB in Higher Dimensions}

In this section, we construct the FUPB in higher dimensions. We
start from a simple observation.

\bex {\rm
 \label{ex:5x5FUPB}
Using Example \ref{ex:4x4FUPB}, we can easily construct an FUPB
$S_1$ in $\we^2\bC^5$. The set $S_1$ consists of the five states in
the set $S$ in Example \ref{ex:4x4FUPB} and four states
$\wetw{5}{i},i=1,2,3,4$. Evidently these nine decomposable states
are orthogonal. By Example \ref{ex:4x4FUPB}, the state orthogonal to
these decomposable states is entangled. So $S_1$ is an FUPB.
 \qed
 }
 \eex

Using a similar method we can indeed construct a nontrivial FUPB
spanning a hyperplane of $\we^2\bC^M$ for any $M\ge4$. So such an
FUPB has cardinality $\binom{M}{2}-1$ and its vectors span a
hyperplane of $\we^2\bC^M$. However, this construction is far from
constructing the FUPB with the least possible cardinality of
$2(M-2)+1=2M-3$ by Lemma~\ref{le:FCES}. Below we construct another
FUPB with smaller cardinality. Let $\ket{a_i}\in\bC^{d_i}$ and
$\ket{b_i}\in\bC^{d_i},i=1,2$. Hence the two product states
$\ket{a_1,a_2}$ and $\ket{b_1,b_2}\in\bC^{d_1d_2}$. The decomposable
state $\wetw{a_1,a_2}{b_1,b_2}$ is a 2-vector in
$\we^2\bC^{d_1d_2}$.

 \bpp
 \label{pp:FUPB,bi}
Let $V$ be a finite-dimensional Hilbert space and let
$V=\oplus_{i=1}^d V_i$ be an orthogonal decomposition.
For $1\le i<j\le d$ let $f_{ij}:V_i\otimes V_j\to\we^2 V$
be the isometric embedding sending
$x_i\otimes x_j\to x_i\we x_j$ for $x_i\in V_i$ and
$x_j\in V_j$.
Suppose that $X_i\su\we^2 V_i$, $i=1,\ldots,d$, are FUPB
and $Y_{jk}\su V_j\otimes V_k$, $1\le j<k\le d$, are UPB.
Then the union, $Z$, of all $X_i$ and all $f_{jk}(Y_{jk})$
is an FUPB in $\we^2 V$.
 \epp
 \bpf
It is obvious that $Z$ is an orthogonal set of bivectors.
Assume that $x\we y\in Z^\perp$ for some $x,y\in V$. Let
$x=\sum x_i$ and $y=\sum y_i$ with $x_i,y_i\in V_i$.
Then
$$
x\we y=\sum x_i\we y_i + \sum_{j<k} (x_j\we y_k-y_j\we x_k).
$$
By the assumption we have $x_i\we y_i \perp X_i$ and $(x_j\otimes
y_k-y_j\otimes x_k) \perp Y_{jk}$, and consequently we must have
$x_i\we y_i=0$ for all $i$ and $x_j\otimes y_k-y_j\otimes x_k=0$ for
$j<k$. (Note that for $a,u\in V_j$ and $b,v\in V_k$ we have
$\braket{u\otimes v}{a\otimes b}=\braket{u\we v}{a\we b}$.) It
follows that $x\we y=0$, and so $Z$ is an FUPB.
 \epf

\begin{example}
 {\rm
As the simplest example for Proposition~\ref{pp:FUPB,bi}, we choose
$d=2$ and $\dim V_1=\dim V_2=3$. It is known the only FUPB in $\we^2\bC^3$ is
the trivial FUPB. So we that the cardinality of $X_1$ and $X_2$ are both $3$, and the cardinality of $Y_{12}$ is $5$, where the
latter is due to the UPB constructed by pentagon~\cite{BDM+99}.
Therefore, by Proposition~\ref{pp:FUPB,bi} we have constructed an
FUPB of cardinality eleven in $\we^2\bC^6$. They span a subspace
whose orthogonal complement is completely entangled and
$4$-dimensional. \qed
 }
\end{example}

%Note that we can
%always choose $S_j$ as the trivial FUPB consisting of
%$\wetw{i}{j},i>j$ and $i,j=\{1,\cdots,d\}$ in $\we^2\bC^d$. So
%Proposition \ref{pp:FUPB,bi} is operational for constructing FUPBs
%of high dimensions. In fact we have

By respectively replacing in this example the trivial FUPB and the
two-qutrit UPB by the FUPB from the sets $S_1,\cdots,S_D$ and a
nontrivial UPB in high dimension, we can construct a nontrivial FUPB
$S$ in the space $\we^2\bC^M$ for any $M\ge4$.  By Example
\ref{ex:5x5FUPB} we may choose the set $S_j$ with
$n_j<{d_j\choose2}$. So $\abs{S}$ is smaller than the cardinality
${M \choose 2}-1$ with which the FUPB is constructed by Example
\ref{ex:5x5FUPB}. However it is still bigger than the least possible
value $2M-3$. Below we construct a nontrivial FUPB with the upper
bound of cardinality. It turns out that this FUPB span a hyperplane
of $\we^N\bC^M$, so it provides the maximum size of FUPB that does
not span $\we^N \bC^M$.

 \bl
 \label{le:maximumFUPB}
There is an FUPB $S$ in $\we^N \bC^M$ such that $\abs{S}={M \choose N}-1$.
 \el
 \bpf
We concretely construct an FUPB $S$ with cardinality ${M \choose
N}-1$. Let $T$ consist of the Slater determinants except the
following six elements: $e_{ij}\we e_{5,\cdots,N+2}$ for
$i,j=1,\cdots,4$ and $i>j$. Suppose $S$ consists of $T$ and the five
states $\ket{a_i\we b_i}\we e_{5,\cdots,N+2}$, $i=1,\cdots,5$ where
the $\ket{a_i}\we \ket{b_i}$ from Example \ref{ex:4x4FUPB} form the FUPB
in $\we^2\bC^4$. So $|S|={M \choose N}-1$ and the only state
orthogonal to the states in $S$ is $\ket{\ph}\we e_{5,\cdots,N+2}$,
where $\ket{\ph}\in\we^2\bC^4$ is orthogonal to the hyperplane
spanned by the $\ket{a_i\we b_i}$. So $\ket{\ph}\we
e_{5,\cdots,N+2}$ is not decomposable, and $S$ is an FUPB. This
completes the proof.
 \epf

In the lemma the state orthogonal to the subspace spanned by $S$ is
entangled. In contrast, any hyperplane in $\we^2\bC^M$ orthogonal to
an entangled state is spanned by a generalized FUPB, see the
paragraph below Proposition \ref{pp:smallest}.

Note that if the set of decomposable vectors
$\ket{\a_1},\cdots,\ket{\a_n}$ is an FUPB in $\we^N \bC^M$, then the
set of decomposable vectors
$\star\ket{\a_1},\cdots,\star\ket{\a_n}$, where $\star$ denotes the
Hodge dual \cite{Sus13}, is an FUPB in $\we^{M-N}
\bC^M$. Using this fact and Example \ref{ex:5x5FUPB}, we may
construct an FUPB of cardinality nine in $\we^3\bC^5$. This is
indeed the simplest tripartite subspace in which a nontrivial FUPB
exists. On the other hand the least possible cardinality of FUPB in
$\we^3\bC^5$ is seven, and we do not have a constructive example of
FUPB reaching this cardinality.

\section{Summary and discussion}

We have studied the FUPB and generalized FUPB in the antisymmetric
subspace $\we^ N \bC^M$, using Slater determinants as an analogue of
product states. First, we have constructed an explicit example of
generalized FUPB of minimum cardinality $N(M-N)+1$ for any
$N\ge2,M\ge4$. The counterpart of the generalized UPB in
$\cH=\bC^{d_1}\ox\cdots\ox\bC^{d_N}$ is known to be $\sum^N_{i=1}
d_i - N + 1$. Second, we have shown that any bipartite antisymmetric
space of codimension two is spanned by decomposable 2-vectors.
Whether this observation holds for $N>2$ remains an open problem. We
also have constructed a subspace in $\we^ N \bC^M$ of codimension
three not spanned by decomposable vectors. Third, we proposed a few
properties of FUPB for $N=2,M=4$ of cardinality five, which is the
FUPB of minimum dimension and cardinality. Using it we have shown
that there is no real FUPB and constructed an example of complex
FUPB of $N=2,M=4$. It is an interesting question to characterize the
canonical form of this simplest FUPB and propose an application to
quantum information theory. In contrast, it is known that the
$3\times3$ UPB is the UPB of minimum dimension and cardinality and
has been fully characterized \cite{dms03}. As an application the
$3\times3$ positive-partial-transpose (PPT) entangled states of rank
four have been fully characterized \cite{cd11}. We also have some
evidence supporting the conjecture that every 5-dimensional subspace
in $\we^2\bC^4$ is spanned by an FUPB. Fourth, we have constructed
FUPBs of arbitrary $N,M$ of cardinality bigger than $N(M-N)+1$,
which is the minimum possible cardinality. An open question is to
construct an FUPB of arbitrary $N,M$ and cardinality $N(M-N)+1$. It
may require ideas and techniques out of the scope of this paper.

\section*{Acknowledgements}

We thank Dragomir Z Djokovic for revising Lemma~\ref{cr:>=d-sPRODstate}, Proposition
\ref{pp:smallest}, \ref{pp:FUPB,bi} and many invaluable comments. We thank Zhengfeng Ji for the delightful discussion which eventually leads to Proposition~\ref{pp:FUPB,bi}. JC and BZ are supported by NSERC, CIFAR and NSF of China (Grant
No.61179030). LC is mainly supported by MITACS and NSERC.

\bibliographystyle{unsrt}

\bibliography{FUPB}

\end{document}